\documentclass{icrc}

 \usepackage{times}
 \usepackage{epsfig}
 \usepackage{graphicx} 

\begin{document}

\title{Diffuse $\gamma$-ray line emission from multiple OB Associations in Cygnus}
\author[1]{S. Pl\"uschke}
\author[1]{M. Cervi\~no}
\author[1]{R. Diehl}
\author[1]{K. Kretschmer}
\affil[1]{MPI f. extraterrestrische Physik, Postfach 1312, 85741 Garching, Germany}
\author[2]{D.H. Hartmann}
\affil[2]{Physics \& Astronomy Dep., Clemson Univ., Clemson, SC 29634, USA}
\author[3]{J. Kn\"odlseder}
\affil[3]{CESR, BP 4346, 31028 Toulouse Cedex, France}

\correspondence{S. Pl\"uschke (stp@mpe.mpg.de)}

\firstpage{1}
\pubyear{2001}


\maketitle

\begin{abstract}
The COMPTEL observations of the diffuse galactic 1.809~MeV emission attributed to
the radioactive decay of $^{26}$Al have confirmed the diffuse nature of this
interstellar emission line. One of the most significant features of the reconstructed
intensity pattern is a flux enhancement in the direction of the Cygnus region. This
region is fairly young and contains a wealth of massive stars, most of them grouped in the
Cygnus OB associations. Multi-frequency model fitting strongly supports the hypothesis of
massive stars and their descendent supernovae being the dominant sources of interstellar
$^{26}$Al as observed by COMPTEL.\\
Massive stars and supernovae are known to impart a large amount of kinetic energy
into the surrounding ISM which lead to shockregions and large cavities. In addition, a large
fraction of the electro-magnetic radiation of massive stars lies in the extreme
ultra-violet causing photoionisation of the surrounding interstellar medium.\\
We applied a population synthesis model in combination with an analytic model
of the expansion of superbubbles to the Cygnus OB associations. The model predicts the
expected 1.809~MeV flux and the $\gamma$-ray line intensity due to interstellar $^{60}$Fe.
We compute the sizes and expansion parameters of the expected HI-structures and the free-free 
emission intensities due to the photoionizing radiation from massive stars within this
region of the sky. We discuss our present understanding of the Cygnus region with respect
to the massive star census and history.\\
Our model assigns about 70\% of the 1.809~MeV intensity to six known OB associations, about
20\% to known isolated sources and roughly 10\% to an unkown diffuse component.
\end{abstract}
\section{Introduction}
\begin{figure}[h!]
 \begin{center}
  \epsfig{figure=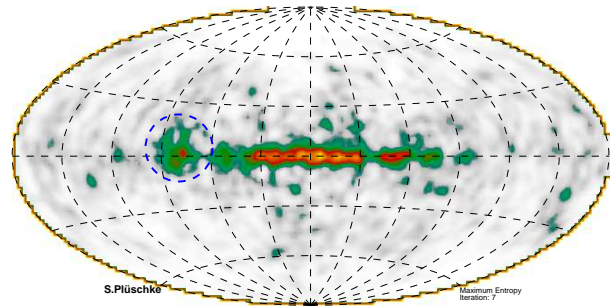,width=8cm,clip=}
   \caption{COMPTEL 1.809~MeV all sky map based on a maximum entropy deconvolution of the
           complete mission data \citep{pl01a}; Cygnus encircled by the ellipse}
  \label{fig:map}
 \end{center}
\end{figure}

The COMPTEL telescope aboard the {\it Compton Gamma-Ray Observatory} has produced a nine
year all sky survey in the MeV regime \citep{sch00,pl01a}. Among these we obtained the all
sky map in the narrow band centered on 1.809~MeV selecting the radioactive decay line
from $^{26}$Al (Fig. \ref{fig:map}). With a mean lifetime of roughly 1~Myr there are many
individually unresolved sources which contribute to the overall diffuse glow of the Galaxy
in this particular $\gamma$-ray line \cite[for a recent review see][]{pd96}.
\\
The bulk of the observed $^{26}$Al flux is due to ejecta from winds of massive stars
and core-collapse supernovae, with relative contributions of 60\% and 40\%,
respectively \citep{kn99}. Classical Novae and AGB stars also contribute to the overall
$^{26}$Al content of the interstellar medium, but certainly to a much lesser extent.
Population synthesis studies and imaging analysis of the 1.809~MeV emission suggest a
total $^{26}$Al mass in the galactic disk of roughly 2~M$_{\odot}$ \citep{kn95}.\\
Multi-wavelength observations of the Milky Way and external galaxies clearly show that
star formation exhibits a strong hierarchical pattern. Most stars form in groups,
associations, clusters, and even larger conglomerates \citep{ee96}. Thus, star fromation
is a strongly correlated process. The transformation of some $10^4$~M$_{\odot}$ of gas
into stars (with typical efficiencies of $\le10\%$) leads to a subsequent dynamic epoch
due to winds and explosions from several hundred stars in the mass range between 8 and,
say, 120~M$_{\odot}$. The kinetic energy input into the surrounding interstellar medium
drives an expanding superbubble filled with hot, tenuous gas into the gaseous disk.
These "super-remnants" can even blow "holes" into the galactic gas disk. There is
much observational evidence for such superbubbles in the Milky Way as well as external
galaxies. Superbubbles provide the means by which the disk and the halo are chemo-dynamicalyl
coupled. In any case, the expansion of these superbubbles provide a mechanism for
distributing freshly synthesized matter such as $^{26}$Al or $^{60}$Fe into the
ambient medium of OB associations. In this way radioactive nuclei may be transported
substantial distances of order $\sim$100~pc before their decay.\\
The Cygnus regions seems to be well suited for a detailed study of such a region. First,
the region contains up to nine OB associations of young to intermediate age testifying a
rather strong star formation activity during the last millions of years. Second, the
Cygnus regions appears to be the strong\-est and most significant emission feature despite
the inner galaxy in the 1.809~MeV map (see Fig. \ref{fig:map}).

\section{Modelling Cygnus}
\begin{figure}[h!]
 \begin{center}
  \epsfig{figure=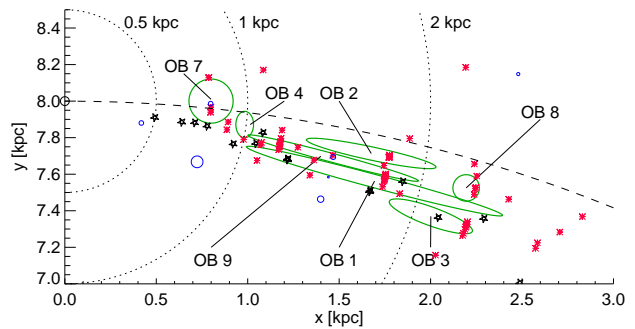,width=8.5cm,clip=}
   \caption{Distribution of O and WR stars (red and black stars) as well as SNR
    (blue circles) and OB associations (green ellipses) in the direction of Cygnus
    as viewed from above the galactic plane. The sun is allocated along the y-axis at
    8.0~kpc.}
  \label{fig:sector}
 \end{center}
\end{figure}

\begin{figure}[h!]
 \begin{center}
  \epsfig{figure=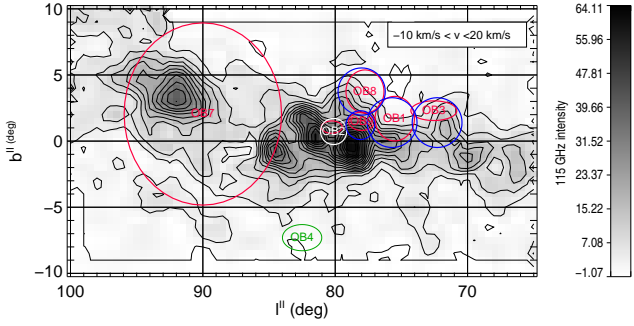,width=8.5cm,clip=}
   \caption{The 115~GHz CO emission towards Cygnus \citep{d01}. The ellipses mark the
    positions of the Cygnus OB associations.}
  \label{fig:co}
 \end{center}
\end{figure}

The Cygnus region as defined by the Cygnus 1.8~MeV emission feature contains numerous
massive stars. The galactic O star catalogue lists 96 O stars in this field \citep{g82}.
In addition, one finds 23 Wolf-Rayet stars in this region of which 14 are of WN-type,
8 of WC-type and one is classified as WO-star \citep{vdh00}. Between 5 and 10 of these
WR stars are believed to be members of OB associations in this area of the sky.
With respect to localized $^{26}$Al production supernova remnants may also be of interest.
The Galactic SNR Catalogue lists 19 remnants in this region, for 9 of those age and distance
have been estimated with sufficient accuracy \citep{g00}. In the subsequently described
model the non-member WR stars and SNRs are taken as stationary isolated sources contributing
some $^{26}$Al to the global content of the Cygnus ISM causing the observed 1.809~MeV emission.
They are modelled with an IMF-weighted average yield assuming a Salpeter-IMF and using the
theoretical yields from \cite{m97} and \cite{ww95}.\\
Beside numerous open clusters the region contains nine OB associations \citep{a70}.
Meanwhile two of those (OB 5 \& 6) are thought to be {\it artificial} association
due to projection and selection effects \citep{gs92}. Furthermore Cygnus OB4 seems to
be poorly determined. Therefore the remaining six associations (Cyg OB1-3 \& OB7-9)
had been chosen as basis for our model. Table \ref{tab:asso} lists the key parameters
of the individual OB associations which are used in our population synthesis model.
\begin{table}[h!]
  \begin{center}
  {\scriptsize
    \begin{tabular}[h]{|c|c c|c|c|c||c|c|} \hline
{\bf Ass.} & {\bf Lon.} & {\bf Lat.} & {\bf Dist.} & {\bf Age} & ${\bf N_{\rm O}}$ & $f_{\rm CO}$ & 
     ${\bf M_{tot}^{[8,120]}}$ \\
     & deg & deg & kpc & Myr & & & ${\rm 10^3 M_{\odot}}$ \\ \hline
1 & 75.5 & +1.7 & 4 - 7 & $1.7\pm0.7$ & 15 & 1.7 & $3.2\pm0.9$ \\ \hline
2 & 80.3 & +0.8 & 2 - 5 & $1.7\pm0.4$ & 40 & 3.0 & $11.3\pm0.7$ \\ \hline
3 & 72.6 & +2.3 & $\le12$ & $1.7\pm0.8$ & 9 & 1.4 & $4.8\pm1.4$ \\ \hline
7 & 90.0 & +2.1 & $\le13$ & $0.8\pm0.1$ & 3 & 2.3 & $2.0\pm0.7$ \\ \hline
8 & 77.8 & +3.8 & $\sim3$ & $2.3\pm0.1$ & 4 & 1.2 & $0.8\pm0.3$ \\ \hline
9 & 78.0 & +1.5 & $<8$ & $1.4\pm0.3$ & 7 & 3.7 & $2.8\pm0.8$ \\ \hline
    \end{tabular}
    \caption{Properties of the Cygnus OB associations; Columns left of the double line
     give the optically determined values ($N_{\rm O}$ gives the number of O stars
     \cite[e.g.][]{a70,gs92,m95}) whereas the columns to the right give model parameters
     as described in the text.}
    \label{tab:asso}}
  \end{center}
\end{table}\\
Figure \ref{fig:sector} shows the distribution of O and WR stars, SNRs and OB associations
in the direction of Cygnus as viewed from above the galactic plane. For SNRs the given
size of the circle represents the appearance at the sky, whereas the ellipses marking
the OB association have a twofold meaning. The minor axis represents the appearance in
a longitude-latitude projection and major axis gives the distance uncertainty for each
association.
\\
Optical population studies towards the Cygnus region are largely hampered by a giant
molecular cloud complex lying at a distance of 500 und 1000~pc \citep{d01}. Figure
\ref{fig:co} shows a velocity integrated intensity image of CO 115~GHz emission towards
the Cygnus region reflecting the large angular extent of the Cygnus Molecular Rift.
By analysing the 2MASS NIR data of a field centered on Cygnus OB2, \cite{kn00} has found
a factor of three more O stars than determined from optical observations. Even more,
Cyg OB2 now is the most massive conglomerate of the stars known in the Milky Way and
resembles more like a young globular cluster than a typical OB association. \cite{kn00}
has found $120\pm20$ O star and $2600\pm400$ OB star members. The slope of the mass
distribution was found to be $\Gamma=1.6\pm0.1$ and total mass of the stellar group
is of the order of $10^5$~M$_{\odot}$. The overall large differences of the inferred
properties of Cygnus OB2 depending more or less on the impact of the molecular clouds
infront of the association let one expect similar effects for the other OB associations
found in Cygnus. In addition, our population synthesis model fails to reproduce the observed
1.809~MeV flux by a factor 2 to 3 when the optically determined properties are directly used as
constrains. We therefore used the CO map, assuming the molecular cloud to lay entirely between
the observer and the association, to calculate the population correction factors for the given
fields in Fig. \ref{fig:co}. The algorithm was calibrated to match the Cygnus OB2 observations
in the optical and NIR regimes. The resulting correction factors $f_{\rm CO}$ are given in
table \ref{tab:asso}.
\\
Only in the case of Cygnus OB9 the derived correction factor is larger than the one
found observationally for Cygnus OB2. In any case these numbers can be thought as prediction
of our model and could be tested by detailed analysis of 2MASS data for the complete
Cygnus region.\\
By using the observed numbers of O stars $N_{\rm O}$ multiplied by the correction factors as
constrains we have generated synthetic populations for each Cygnus association by means of
a Monte Carlo sampling of the initial mass function. These populations are statistically
equivalent to the real populations and their fluctuations represent the uncertainties
due to the a priori unkown real distribution of stellar masses in the Cygnus associations.
By means of our population synthesis model \cite[e.g.][]{pl99,pl00} we computed the time
evolution of the interstellar masses of $^{26}$Al and $^{60}$Fe, the kinetic power due
to stellar winds and supernova explosions and the emission of ionizing photons for
each synthetic population. In addition, we also calculate the number of WR stars among the
population members. By adding these numbers from each Cygnus association one finds an expected
number of WR stars being observable as members today of 7 to 12. This number must be compared
to the number of WR stars being labelled as association members in the Galactic WR Star Catalogue
of 5 to 10 \citep{vdh00}. This already proves the consistency of our model.\\
The time profiles of the interstellar masses of radioactive nuclei can easily be converted into
lightcurves in the appropriate $\gamma$-ray lines due to the decay of these isotopes. Beside the
statistic and systematic uncertainties of the population description \citep{pl00} these lightcurves 
are additionally hampered by the sometimes distance uncertainties of the associations.
\begin{figure}[h!]
 \begin{center}
  \epsfig{figure=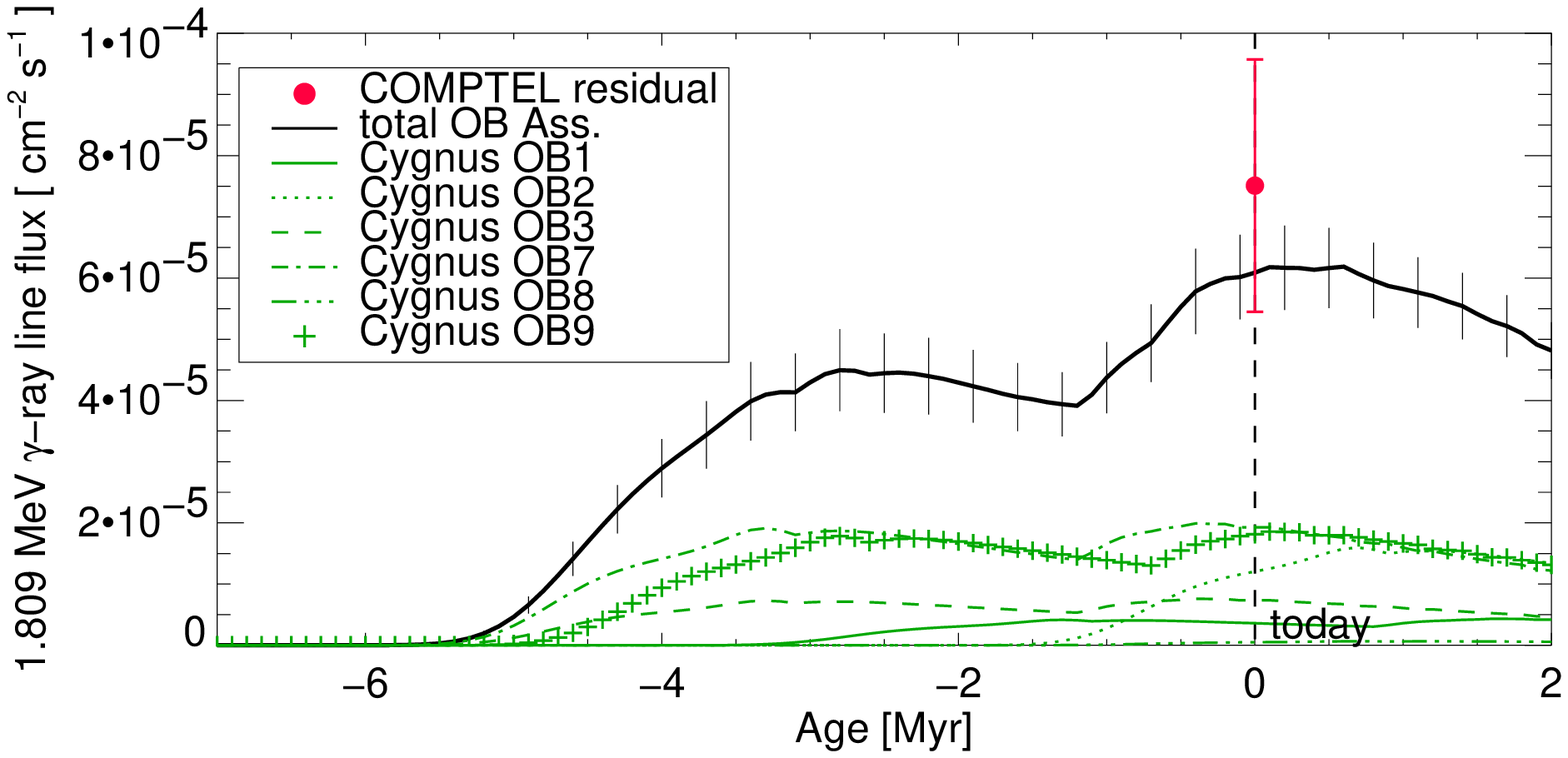,width=8cm,clip=}
  \epsfig{figure=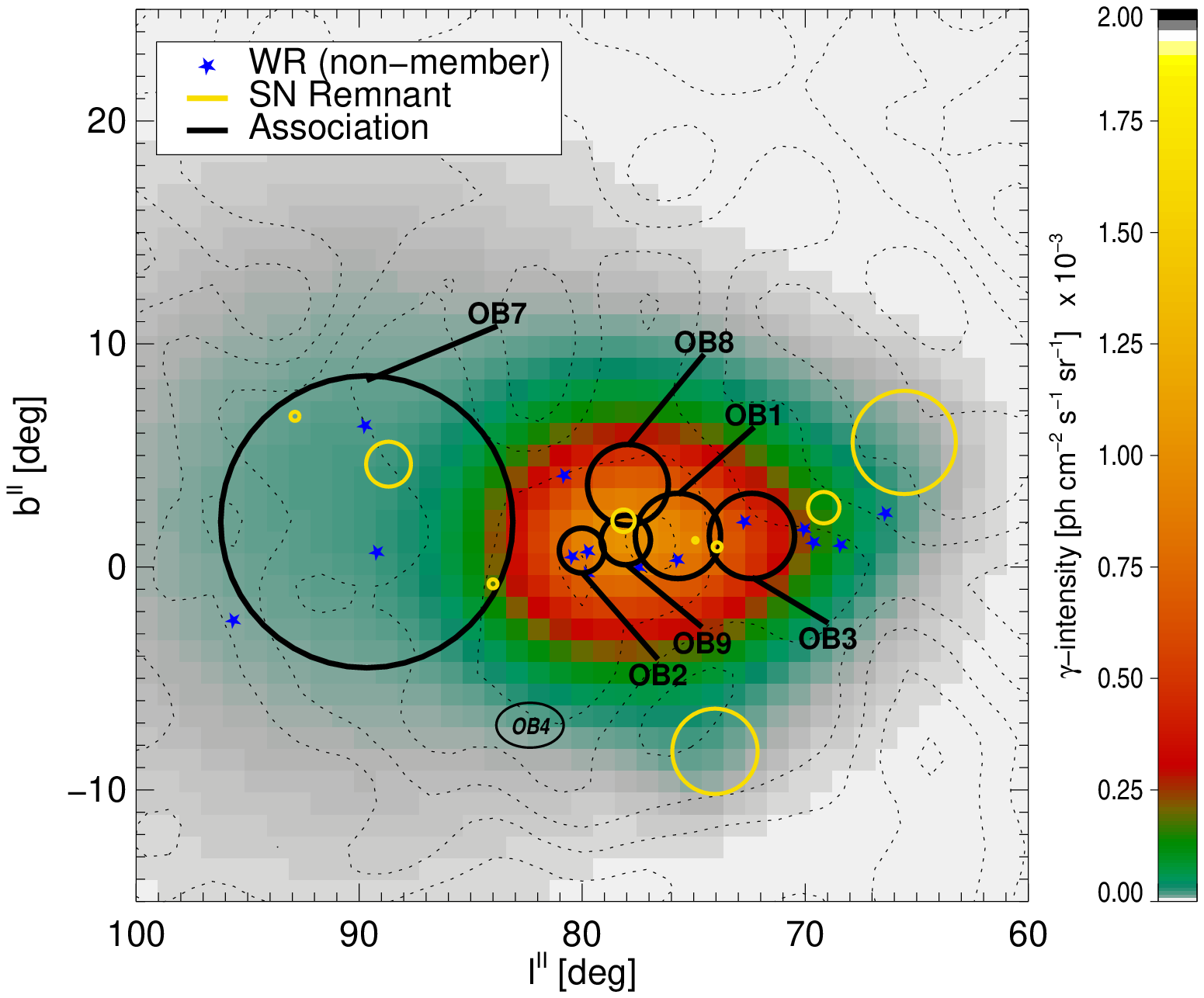,width=8.5cm,clip=}
   \caption{1.809~MeV emission model of the Cygnus region:
    (upper) The diagram shows the combined lightcurve of the expected 1.809~MeV flux and
    the contributions of individual OB associations. The given COMPTEL value is difference
    of the real observed flux and the expected contribution from isolated sources.
    (lower) assuming a mean ambient density of $10^3\,{\rm cm^{-3}}$ the $^{26}$Al
    is imparted to the ISM by OB associations and non-member WR stars and SNR; the
    intensity distribution is smoothed with a Gaussian of $3.8^{\circ}$ (FWHM). The
    dotted contours represent the COMPTEL 1.8~MeV image of this region.}
  \label{fig:model}
 \end{center}
\end{figure}\\
The upper panel of Fig. \ref{fig:model} shows the expected lightcurves for the individual
and the integrated contribution of the Cygnus OB associations to the observed 1.809~MeV intensity.
The COMPTEL data point gives the observed flux minus the contribution of the stationary,
isolated sources. The extinction corrected population synthesis model is therefore consistent
with this residual and on this basis we can argue that about 70\% of the observed 1.809~MeV
flux is due to OB associations in Cygnus whereas approximately 20\% can be attributed to
known single sources.\\
The release of freshly synthesized radionuclids is directly coupled into the injection of kinetic
energy to the interstellar medium. Typical stellar wind power is of the order of $10^{37}\,{\rm
erg\,s^{-1}}$ whereas supernovae are believed to impart $\sim10^{51}\,{\rm erg}$ of kinetic energy 
to the surrounding medium \cite[but see][]{t98}. Therefore OB associations are believed to drive large
cavities into the interstellar medium which are filled with a hot, tenuous medium. These super-remnants
can reach linear dimensions of several 100~pc and are often called superbubbles. Because of the high
ejection velocities (several 1000~km/s) and the low density interior one can expect the freshly ejected
radionuclids to travel large distance before their decay. Indeed, assuming a mean velocity of
1000~km/s $^{26}$Al nuclei could travel distances of 200~pc within one fith of their mean lifetime.
Therefore the diffuse nature of the observed 1.809~MeV emission has a twofold origin. First
due to the long lifetime many sources contribute to a diffuse glow and second the effects of
the dynamical ejection tend to smear out the distribution of parent nuclei.\\
We used a 1D analytical prescription of expanding superbubbles to model these effects \citep{pl01b}.
This model could be at large compared with the model described by \cite{ss95}. This model assumes
a uniform ambient density and drives a thin shell blown by a variable energy source. The energy
losses due to radiation are taken properly into account. By taking the time profiles of the
wind power and cumulative mass loss for each synthetic association as driver we computed
the evolution of the bubble radii. In the final step we distributed the expected mass of
$^{26}$Al coming from the individual associations uniformly over a circle at the sky
representing the mean expected bubble radius and do a quantitative model comparison with
the observed 1.809~MeV data by means of Maximum Likelihood fitting. The lower panel in Fig.
\ref{fig:model} shows the resulting $\gamma$-ray intensity distribution of a model assuming
a mean ambient density of $10\,{\rm cm^{-3}}$. In addition, this image has been convolved with a
Gaussian of $3.8^{\circ}$ (FWHM) to resemble the COMPTEL imaging capabilities. The model fits
prefer a model with mean ambient density between 20 and $30\,{\rm cm^{-3}}$, which is consistent
with the ambient densities inferred by others \cite[e.g.][]{c94}.\\
In addition, the model fitting supports the idea of a low-intensity component with an extent as
large as the total Cyg\-nus region. The most plausible origin for this might be a foreground source.
Nevertheless, contributions from novae and/or AGB stars, leading in general to a much smoother,
low-intensity emission, can not be excluded. This feature may also be seen as hint in favour of
a real existence of the association Cygnus OB6 \&6, which indeed should have very large angular
extents \citep{a70}.\\
Finally, from this model of the massive star population of the Cygnus region a prediction of the
$\gamma$-ray line flux due to radioactive decay of $^{60}$Fe emerges. Within the given framework
we expect a total flux $(8.6\pm5.1)\cdot10^{-6}\,{\rm ph\,cm^{-2}\,s^{-1}}$, which is well
below the COMPTEL sensitivity limit and even for INTEGRAL we do not expect much to be seen.

\section{Summary}
We have shown that the extented 1.809~MeV line flux in the Cygnus star forming region can be
consistently modelled with the $^{26}$Al nucleosynthesis in massive stars and its ejection
by stellar winds and core-collapse supernovae. Based on known properties of the population of
massive stars in Cygnus we construct a time dependent 1.809~MeV light curve. We find the
optically determined number of massive stars is insufficient to match the present day flux
level. With an extinction correction based on detailed CO maps, however, the $\gamma$-ray flux
can be reproduced. This implies an increased number of massive stars in the individual OB
associations above those deduced from optical surveys, which is confirmed by a detailed analysis
of the 2MASS NIR data of the Cygnus OB2 region \citep{kn00}.We estimate relative contributions
from OB associations and isolated WR stars and SNRs of 70\% and 20\%, respectively. Another 10\%
of the observed 1.809~MeV emission might be due to a low-level foreground.\\ \\
Our study support the basic notion of dynamic injection and distribution of freshly synthesized
$^{26}$Al. Star formation correlated in space and time leads to expanding superbubbles filled with
new radioactivity. Within the approximation of spherically symmetric thin shells expanding in a
uniform ambient medium we estimate an exterior density between 20 and $30\,{\rm cm^{-3}}$.


\vspace{-0.5cm}
\begin{thebibliography}{99}

\bibitem[Alter et al.(1970)]{a70}
Alter, J. et al., {\it Catalogue of star clusters and associations},
Budapest: Akademiai Kiado, 1970.

\bibitem[Comeron \& Torra(1994)]{c94}
Comeron, F. and Torra., J., {\it ApJ} {\bf 423}:652+, 1994.

\bibitem[Dame et al.(2001)]{d01}
Dame, T.M. et al., {\it ApJ} {\bf 547}:792-813, 2001.

\bibitem[Elmegreen \& Efremov(1996)]{ee96}
Elmegreen, B.G. and Efremov, Y.N., {\it ApJ} {\bf 466}:802+, 1996.

\bibitem[Garmany et al.(1982)]{g82}
Garmany, C.D. et al., {\it ApJ} {\bf 263}:777+, 1982.

\bibitem[Garmany \& Stencel(1992)]{gs92}
Garmany, C.D. and Stencel, R.E., {\it A\&AS} {\bf 94}:211-244, 1992.

\bibitem[Green(2000)]{g00}
Green, D.A., {\it Catalogue of galactic SNRs (August 2000)},
URL: {\it http://www.mrao.cam.ac.uk/surveyz/snrs/}, 2000.

\bibitem[Kn\"odlseder et al.(1999a)]{kn95}
Kn\"odlseder, J. et al., {\it A\&A} {\bf 344}:68-82, 1999.

\bibitem[Kn\"odl\-seder(1999b)]{kn99}
Kn\"odlseder, J., {\it ApJ} {\bf 510}:915-929, 1999.

\bibitem[Kn\"odlseder(2000)]{kn00}
Kn\"odlseder, J., {\it A\&A} {\bf 360}:539-548, 2000.

\bibitem[Massey et al.(1995)]{m95}
Massey, P. et al., {\it ApJ} {\bf 454}:151+, 1995.

\bibitem[Meynet et al.(1997)]{m97}
Meynet, G. et al., {\it A\&A} {\bf 320}:460-468, 1997.

\bibitem[Pl\"uschke et al.(1999)]{pl99}
Pl\"uschke, S. et al., in {\it Astronomy with Radioactivities}, Ringberg,
Sep. 1999, eds. R. Diehl \& D. Hartmann, MPE Rep. 274, 1999.

\bibitem[Pl\"uschke et al.(2000)]{pl00}
Pl\"uschke, S. et al., in {\it The Influence of Binaries on Population Studies}, Brussels,
Aug. 2000, eds. W. van Rensbergen, D. Vanbeveren \& B. de Loore, KAP (in press), 2000.

\bibitem[Pl\"uschke et al.(2001a)]{pl01a}
Pl\"uschke, S. et al., {\it Proc. of the $4^{th}$ INTEGRAL Workshop}, Alicante/Spain,
ESA-SP series (in press), 2001.

\bibitem[Pl\"uschke(2001b)]{pl01b}
Pl\"uschke, S., PhD thesis, TU M\"unchen, 2001.

\bibitem[Prantzos \& Diehl(1996)]{pd96}
Prantzos, N. and Diehl, R., {\it Phys. Rep.} {\bf 267(1)}:1-70, 1996.

\bibitem[Sch\"onfelder et al.(2000)]{sch00}
Sch\"onfelder, V. et al., {\it A\&AS} {\bf 143}:145-179, 2000.

\bibitem[Shull \& Saken(1995)]{ss95}
Shull, M.J. and Saken, {\it ApJ} {\bf 444}:663-671, 1995.

\bibitem[Thornton et al.(1998)]{t98}
Thornton, K. et al., {\it ApJ} {\bf 500}:95+ ,1998.

\bibitem[van der Hucht et~al.(2001)]{vdh00}
van der Hucht K., et~al., {\it New Astr. Rev.} {\bf 45(3)}:135-232, 2001. 

\bibitem[Woosley \& Weaver(1995)]{ww95}
Woosley, S.E. and Weaver, {\it ApJS} {\bf 101}:181+, 1995.

\end{thebibliography}
\end{document}